\documentstyle[aps,prl,citesort,epsf]{revtex}
\newcommand{\eqref}[1]{Eq.~(\protect\ref{#1})}
\newcommand{\figref}[1]{Fig.~\protect\ref{#1}}

\begin{document}

\draft
\twocolumn

\title{Comment on ``Optimal Periodic Orbits of Chaotic Systems''}

\author{
Scott~M.\ Zoldi\cite{CNCS-address}\cite{zoldi-email} and
Henry~S.\ Greenside\cite{CNCS-address}
}

\address{
Department of Physics\\
Duke University, Durham, NC 27708-0305
}

\date{June 25, 1997}

\maketitle

\vspace{.3in}

In a recent Letter \cite{Hunt96prl} (see also
\cite{Hunt96pre}), Hunt and Ott argued that {\em
short}-period unstable periodic orbits (UPOs) would be the
invariant sets associated with a chaotic attractor that are
most likely to optimize the time average of some smooth
scalar performance function. In this Comment, we show that
their conclusion does not hold generally and that optimal
time averages may specifically require long-period
UPOs. This situation can arise when long-period UPOs are
able to spend substantial amounts of time in a region of
phase space that is close to large values of the performance
function.

Our counterexample is based on a numerical study of the
Lorenz equations~\cite{Lorenz63} for the classical parameter
values $r=28$, $\sigma=10$, and $b=8/3$, for which most
initial conditions evolve towards the familiar
butterfly-shaped chaotic attractor. We chose as a
physically-motivated performance function the Nusselt
number~$N(t)$ which is a dimensionless spatially-averaged
measure of the vertical heat transport of a convection
cell~\cite{Busse78}; for the Lorenz equations, it is given
in terms of the Lorenz variables~$x$ and~$y$ by
\begin{equation}
N = 1 + { 2 \over b r } x y . \label{nusselt-number}
\end{equation}
We evaluated the time average of this expression, $\langle N
\rangle_t$, for each of 718~distinct UPOs of the Lorenz
equations which had been calculated by a damped-Newton
numerical method described elsewhere~\cite{Zoldi97KSUPOS}. A
comparison of the calculated UPOs with a known binary
symbolic dynamics showed that all UPOs of period $T \le 6.0$
were found by our method.

Our results are summarized in \figref{nusselt-vs-period}. Panel~(a)
shows how the time-averaged Nusselt number $\langle N \rangle_t$
depends on the period~$T$ of the UPOs. The shortest period UPOs give
average heat transport values that are less than the average value for
the chaotic attractor itself (indicated by the horizontal line
at~$N=2.68$). The largest values of $\langle N \rangle_t$ occur with
{\em increasing} period. All these values should be compared with the still
larger value $N=2.93$ that is attained for the two nonzero unstable
fixed points $x=y=\pm \sqrt{b(r-1)}, z=r-1$.  Panel~(b) shows that the
long-period large-Nusselt-number UPOs achieve a larger value
of~$\langle N \rangle_t$ specifically because they are asymmetric and
spend more time in the vicinity of a nonzero fixed point. Symbolic
dynamics predicts the existence of these asymmetric UPOs up to an
arbitrarily long period. Since the fixed points lie distinctly outside 
the Lorenz attractor, the Nusselt number of all UPOs should be
strictly bounded by the fixed-point value.

\vspace{0.3in}
\centerline{\epsfysize=3.0in \epsfbox{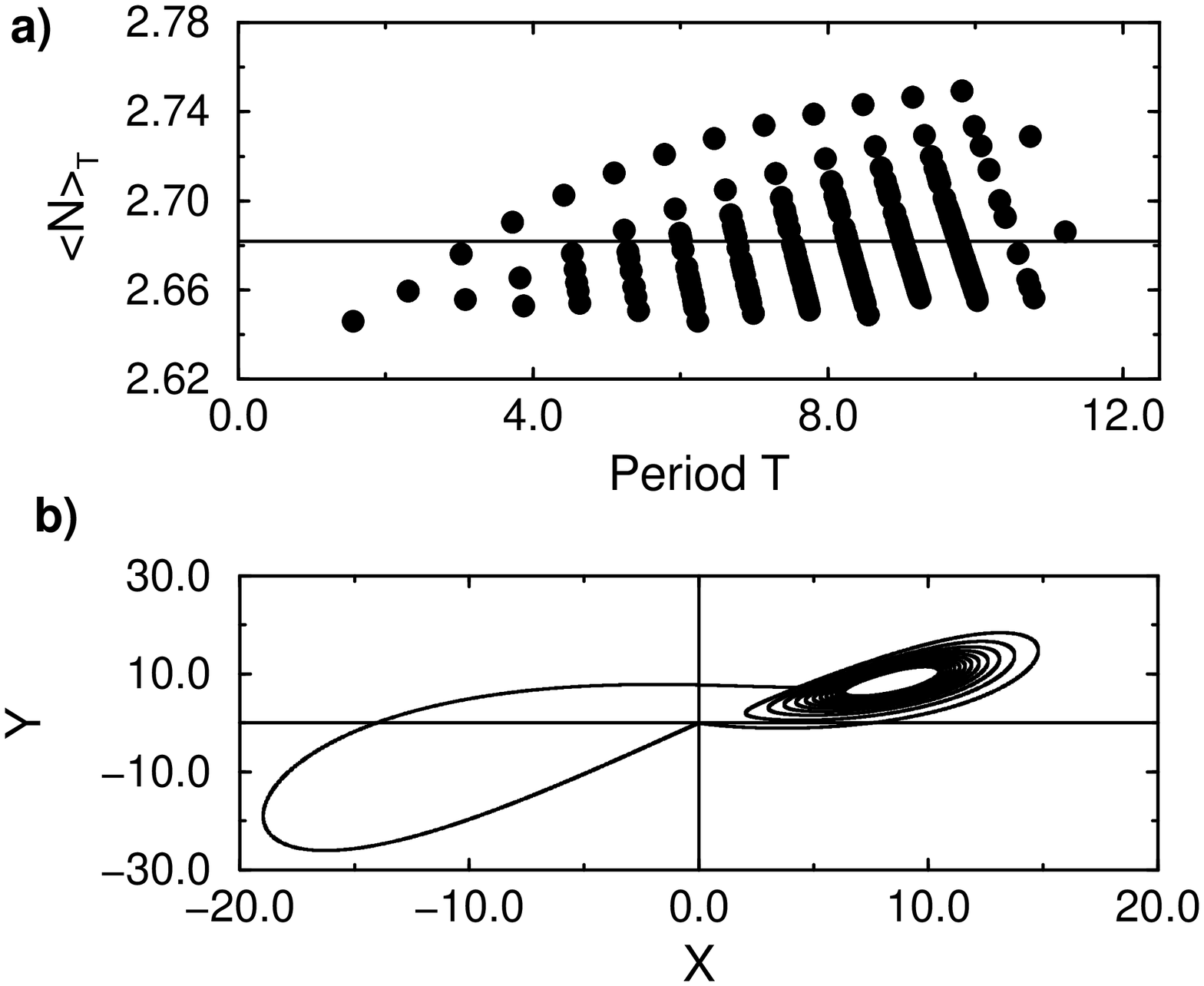}}

\begin{figure}
\caption{ 
{\bf (a)} Time-averaged Nusselt number~$<N>_{t}$
versus period~$T$ for unstable period orbits (UPOs)
associated with the Lorenz attractor. {\bf (b)} Projection
on the $x-y$ plane of a long-period UPO with~$T=9.82$
and~$N=2.750$, showing the asymmetry that allows the UPO to
spend much time near a fixed point with large Nusselt
number.  }
\label{nusselt-vs-period}
\end{figure}

We thank Brian Hunt and Ed Ott for discussions about the
topic of this Comment. This research was supported by a DOE
Computational Science Graduate Fellowship, by NSF grants
NSF-DMS-93-07893 and NSF-CDA-92123483-04, and by DOE grant
DOE-DE-FG05-94ER25214.

% For final version, comment out the two \biblio... lines
% and uncomment the \input{references} lines

%\bibliographystyle{prsty}  % entries in order of citation    
%\bibliography{chaos,control,convection,hsg,lasers,periodic}

\newpage

%\input{figures}
%\centerline{\epsfysize=8.5in \epsfbox{test.eps}}

\end{document}